\documentclass[runningheads]{llncs}
\pdfoutput=1
\usepackage{graphicx}
\usepackage{xcolor}
\usepackage{amsmath}
\graphicspath{{Figures/}}

\usepackage{pdfpages}
\usepackage{hyperref}
\usepackage[absolute]{textpos}
\newcommand{\id}{{\tt id}}
\newcommand{\cnc}{\ensuremath{\parallel}}

\usepackage{hyperref}

\begin{document}
\begin{textblock}{30}(1,0.2)
\noindent\tiny  Preprint version – please cite as: Kermezis G., Limniotis K., Kolokotronis N. (2021) User-Generated Pseudonyms Through Merkle Trees. In: Gruschka N., Antunes L.F.C., \\ Rannenberg K., Drogkaris P. (eds) Privacy Technologies and Policy. APF 2021. Lecture Notes in Computer Science, vol 12703. Springer, Cham. \\ https://doi.org/10.1007/978-3-030-76663-4\_5
\end{textblock}

\title{User-generated pseudonyms\\ through Merkle trees
\thanks{This work was supported by CYBER--TRUST project, which has received funding from the European Union's Horizon 2020 research and innovation programme under grant agreement no. 786698. This work reflects the authors' view and agency is not responsible for any use that may be made of the information it contains.}}

\titlerunning{User-generated pseudonyms through Merkle trees}

\author{Georgios Kermezis\inst{1}\and
Konstantinos Limniotis\inst{1,2}\orcidID{0000-0002-7663-7169} \and
Nicholas Kolokotronis\inst{3}\orcidID{0000-0003-0660-8431}}
\authorrunning{G. Kermezis, et al.}

\institute{School of Pure \& Applied Sciences, Open University of Cyprus, 2220 Latsia, Cyprus\\ 
\texttt{\{\href{mailto:georgios.kermezis@st.ouc.ac.cy}{georgios.kermezis},\,\href{mailto:konstantinos.limniotis@ouc.ac.cy}{konstantinos.limniotis}\}@ouc.ac.cy}
\and
Hellenic Data Protection Authority, Kifissias 1-3, 11523, Athens, Greece\\
\texttt{\href{mailto:klimniotis@dpa.gr}{klimniotis@dpa.gr}}
\and
Department of Informatics and Telecommunications, University of Peloponnese,  Akadimaikou G.K. Vlachou Street, 22131 Tripolis, Greece\\
\texttt{\href{mailto:nkolok@uop.gr}{nkolok@uop.gr}}}

\maketitle

\begin{abstract}
A  pseudonymisation technique based on Merkle trees is described in this paper. More precisely, by exploiting inherent properties of the Merkle trees as cryptographic accumulators, we illustrate how user-generated pseudonyms can be constructed, without the need of a third party. Each such pseudonym, which depends on several user's identifiers, suffices to hide these original identifiers, whilst the unlinkability property between any two different pseudonyms for the same user is retained; at the same time, this pseudonymisation scheme allows the pseudonym owner to easily prove that she owns a pseudonym within a specific context, without revealing information on her original identifiers. Compared to other user-generated pseudonymisation techniques which utilize public key encryption algorithms, the new approach inherits the security properties of a Merkle tree, thus achieving  post-quantum security. 

\end{abstract}

\section{Introduction}
\label{sec:introduction}

Pseudonymisation of personal data constitutes an important privacy enhancing technique that, when appropriately implemented, suffices to provide specific data protection safeguards. More precisely, the data pseudonymisation may give rise to protecting (hiding) the real identities of the individuals (which is related, as a data protection goal, to data confidentiality), as well as to unlinkability of individuals across different application domains. Moreover, pseudonyms  can  also be used in some cases to ensure verification of the actual identity of the individual (which is related, as a data protection goal, to data integrity)  \cite{enisa:a,enisa:b}.  Therefore, taking into account the six data protection goals as they have been presented in \cite{Hansen} for addressing the legal, technical, economic, and societal dimensions of privacy and data protection in complex IT systems - namely confidentiality, integrity, unlinkability, availability, intervenablity and transparency - the pseudonymisation may contribute in ensuring (at least) the three of them.

The aforementioned data protection goals of pseudonymisation are implied in the European General Data Protection Regulation (Regulation (EU) 2016/679 or GDPR). There are several references to pseudonymisation within the GDPR, mainly as the vehicle for providing appropriate data protection safeguards in several cases, such as towards achieving the so-called {\em data protection by design} principle. However, choosing a proper pseudonymisation technique is not always an easy task, since there are different parameters that need to be considered each time, taking into account the specific scenario that the pseudonymisation is to be used \cite{enisa:a,enisa:b}.

One quite challenging use case of pseudonymisation is the one that the user's pseudonym is being generated in the user's environment - i.e. user-generated pseudonyms. In such a scenario,  neither the data controller (as is defined in the GPDR) nor any other (trusted or not) third party is actively employed in the process of deriving the pseudonyms; instead, the individuals by themselves, via a specific process in a decentralised approach, generate pseudonyms which in turn are being subsequently used by data controllers. 
Several such pseudonymisation approaches have been proposed in the literature (e.g. \cite{Schartner,Lehnhardt,Tunaru}). One of the most known scenarios of user-generated pseudonyms is the the case of several blockchain systems (such as the case of Bitcoin), in which the users are being identified by a meaningless identifier (i.e. the pseudonym, being called {\em address} in this case) which is uniquely associated with a relevant cryptographic key corresponding to its owner.

As stated in \cite{Lehnhardt}, when designing such a decentralised approach for pseudonym generation, we are mainly interested in fulfilling the following requirements: i) ease of use, ii) linking a pseudonym to its owning user should not be possible for any other than the user herself, unless it is explicitly permitted, iii) in cases that users may have multiple pseudonyms, it should not be possible to identify different pseudonyms as belonging to the same user, iv) injectivity, in terms that the pseudonym generation process should avoid duplicates, v) flexibility, i.e. it should be possible to add new pseudonyms to the user entities with minimal effort. 

In this paper, we  explore the notion of the so-called cryptographic accumulators, in order to derive user-generated pseudonyms with some specific properties. Cryptographic accumulators are data structures based on cryptographic primitives to efficiently implement set membership operations \cite{Ozcelik}. They allow to accumulate a finite set of values $\{x_1, x_2,\ldots, x_n\}$ into a succinct value $X$. Therefore, they may constitute a convenient tool to derive pseudonyms (corresponding to the relevant succinct values) that are contingent on a set of initial identifiers (corresponding to the relevant set of values), so as to allow extracting the information whether a given identifier corresponds to a given pseudonym. To achieve this goal, we appropriately utilize the Merkle trees \cite{Merkle} - which is a case of a cryptographic accumulator - as the means to provide a new pseudonymisation technique. The generic idea of using Merkle trees for pseudonymisation purposes has been very recently discussed in \cite{enisa:c}. In our approach, we propose a new pseudonymisation scheme such as, for a user $A$ with some domain-specific identifiers $\id_{A_{1}},\ldots,\id_{A_{n}}$, a pseudonym $P_{A}$ of $A$ can be generated by the user $A$ herself, satisfying the following properties:
\begin{description}
    \item[P1.]The pseudonym $P_{A}$ depends on all $\id_{A_{1}},\ldots,\id_{A_{n}}$. 
    \item[P2.]Knowledge of $P_{A}$ does not allow revealing any of the original identifiers $\id_{A_{i}}$, $i=1,2,\ldots,n$ .
    \item[P3.]The user $A$ can prove, whenever she wants, that any $\id_{A_{i}}$, $i=1,2,\ldots,n$,  corresponds to $P_A$, without revealing any other information on the remaining identifiers $\id_{A_{j}}$, $j\in\{1,2,\ldots,n\}\setminus\{i\}$ .
    \item[P4.]The user may generate several such pseudonyms $P_{A}^{(1)},\ldots,P_{A}^{(s)}$, with the above properties, being pairwise unlinkable. 
    \item[P5.]Two different users $A$ and $B$ will always generate different pseudonyms $P_{A}$ and $P_{B}$, regardless the number, the types and the values of their original identifiers. 
\end{description}
Therefore, the new pseudonymisation technique satisfies the properties described in \cite{Lehnhardt} (implementation issues will be subsequently analysed), enriched with some additional properties that may be of high importance in specific scenarios, as discussed next. Actually, due to the property ${\bf P3}$, the user $A$ may prove, if she wants, that two different pseudonyms $P_{A}^{(1)}$ and $P_{A}^{(2)}$, even if they have been given to two different organisations $\mbox{Org}_1$ and $\mbox{Org}_2$ respectively, correspond to her; such a proof of pseudonym's ownership though (i.e. proving to $\mbox{Org}_1$ that she owns $P_{A}^{(2)}$ in $\mbox{Org}_2$ and/or vice versa), does not reveal any additional information on the original identifiers of $A$ to either $\mbox{Org}_1$ or $\mbox{Org}_2$. 

Moreover, it should be pointed out that the cryptographic strength of the above properties are strongly related to the cryptographic strength of the Merkle tree as an one-time signature scheme, which is known to be post-quantum secure under specific assumptions on the underlying hash function \cite{Buchmann,Buchmann:b}  (see subsection \ref{sub:security:analysis}). This is an important property, taking also into account that other known techniques on deriving user-generated pseudonyms (such as the aforementioned techniques in \cite{Schartner,Lehnhardt,Tunaru})) rely on conventional public key cryptographic schemes which are not post-quantum secure.

The rest of the paper is organised as follows. First, the necessary background is given in Section \ref{sec:preliminaries}, covering both the basic elements of the legal framework (subsection \ref{sub:gdpr}) and the typical Merkle trees (subsection \ref{sub:merkle}). Next, the basic idea on the proposed pseudonymisation technique is presented in Section \ref{sec:analysis}; this section also includes a discussion on the security properties and on implementation issues, as well as on possible application scenarios for this technique. Finally, concluding remarks are given in Section \ref{sec:conclusions}.

\section{Preliminaries}
\label{sec:preliminaries}

\subsection{Pseudonymisation and  data protection: legal framework}
\label{sub:gdpr}
The European Regulation (EU) 2016/679 (2016) ---known as the {\em General Data Protection Regulation} or GDPR--- constitutes the main legal instrument for personal data protection in Europe, which applies  to all organizations that process personal data of individuals residing in the European Union, regardless of the organizations' location, which can be outside European Union.

The term {\em personal data} refers to any information relating to an identified or identifiable natural person, that is a person who can be identified (being called {\em data subject}). {\em Personal data processing} means any operation that is performed on personal data, including the collection, recording, structuring, storage, adaptation or alteration, retrieval, use, disclosure by transmission, dissemination, combination and erasure. The GDPR codifies the basic principles that need to be guaranteed when personal data are collected or further processed and sets specific obligations to the {\em data controllers} - i.e. the entities that, alone or jointly with others, determine the purposes and means of the processing of personal data. Amongst them, the so-called {\em data minimisation principle} refers to the necessity that the personal data shall be  
adequate, relevant and limited to what is necessary in relation to the purposes for which they are processed (art. $5$ of the GDPR). 

The data minimisation, as a fundamental principle, spans the entire text of the GDPR: for example, is it explicitly mentioned in art. $25$ towards ensuring the 
{\em data protection by design} principle, which in turn constitutes an important challenge involving various technological and organisational aspects \cite{Alshammari}. Moreover, the art. $11$ of the GDPR states  that if the purposes for which the data controller processes personal data  do not or do no longer require the identification of an individual, then the controller shall not be obliged to maintain, acquire or process additional information in order to identify the data subject.

In light of the above requirements, data pseudonymisation plays an important role in data protection. From an engineering perspective, a pseudonym
is defined as an identifier of a subject, which is different from the subject's {\em real name} \cite{Pfitzmann,Akil}, whereas the types of pseudonyms  may be distinguished by the context of use \cite{Pfitzmann}. Typically, a pseudonym replaces a data subject's identifier, with the latter one being able to explicitly identify the data subject within a specific context; for example, the original identifier can be a combination of first name and last name, an e-mail address, or even a device/network identifier (e.g. an IP address, a device ID etc.) which in turn constitute personal data when the device is associated with an individual (see also \cite{Chatzistefanou}).  

The GDPR also defines pseudonymisation as {\em the processing of personal data in such a manner that the personal data can no longer be attributed to a specific data subject without the use of additional information, provided that such additional information is kept separately and is subject to technical and organisational measures to ensure that the personal data are not attributed to an identified or identifiable natural person}. As the GDPR explicitly states, pseudonymous data are personal and not anonymous data, despite the fact that there is often a confusion in characterizing pseudonymous data as anonymous (see, e.g., \cite{Chatzistefanou} for a discussion on this). However, with a properly implemented pseudonymisation scheme, pseudonymous data should not allow revealing the original identifier without some {\em additional information}; this piece of information can be, for example, a cryptographic key which is protected - and that's why a pseudonymisation technique often relies on utilizing a cryptographic function to identifiers or other identity-related information (see, e.g., \cite{enisa:a,enisa:b} and the references therein).

\subsection{Merkle trees}
\label{sub:merkle}
\begin{figure}
\includegraphics[width=1\textwidth]{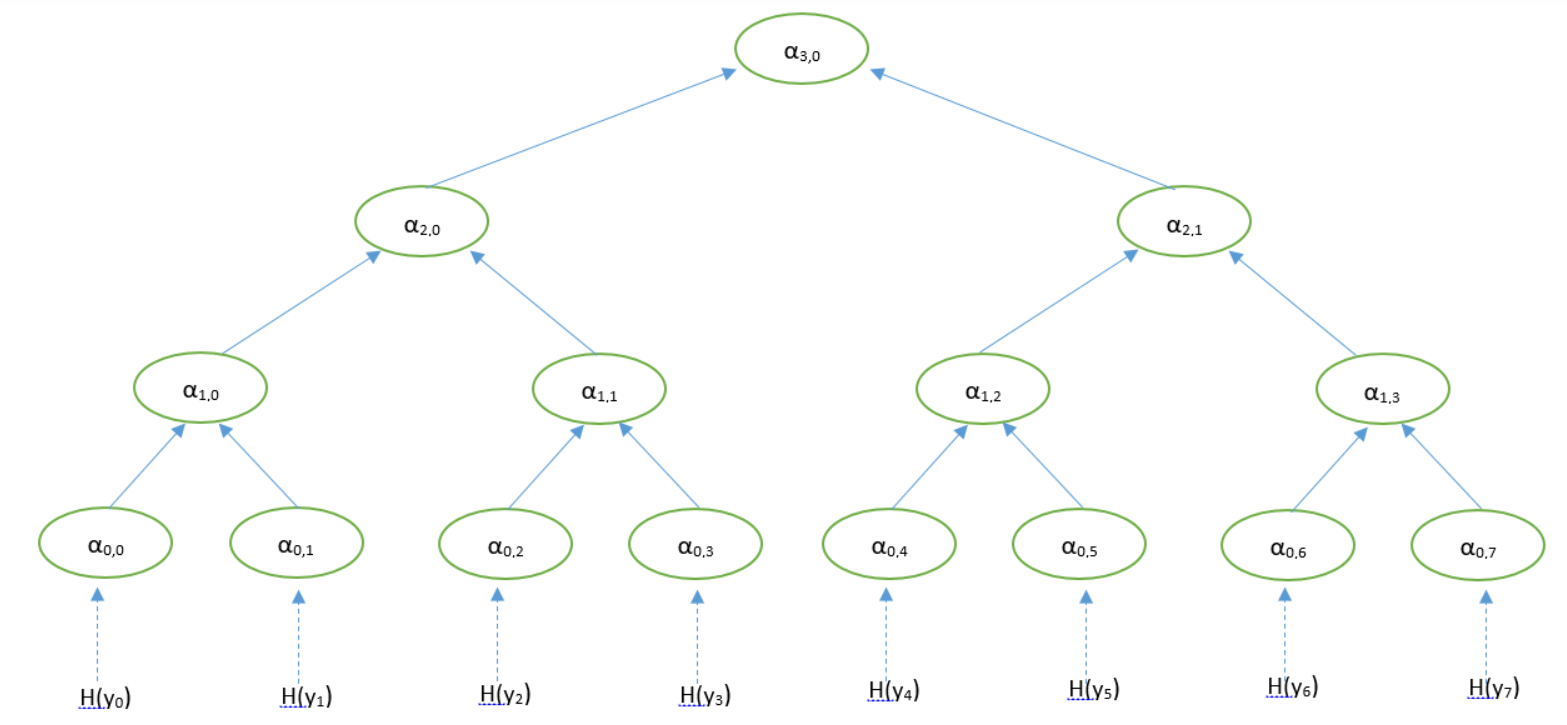}\centering
\caption{A classical Merkle tree with $2^3$ leaves.} 
\label{fig:merkle}
\end{figure}

A Merkle tree is a binary tree based on a cryptographic hash function $H$. Having as starting point $N$ values $y_{0},\ldots, y_{N-1}$, where $N=2^n$ for some integer $n$, the $i$-th leaf node is labeled with the corresponding hash value $H(y_{i})$ of $y_i$, whereas every inner node is labeled with the hash value formed from the concatenation of its children's labels. The label of the root node is the accumulated value, which is clearly contingent on all $y_{0},\ldots, y_{N-1}$. For example, in Fig. \ref{fig:merkle} which illustrates a Merkle tree of $2^3=8$ leaves (i.e. of height $3$), it holds $a_{1,0}=H(a_{0,0}\cnc a_{0,1})=H(H(y_{0})\cnc H(y_{1}))$, $a_{2,0}=H(a_{1,0}\cnc a_{1,1})$ and $a_{3,0}=H(a_{2,0}\cnc a_{2,1})$.

\begin{figure}
\includegraphics[width=1\textwidth]{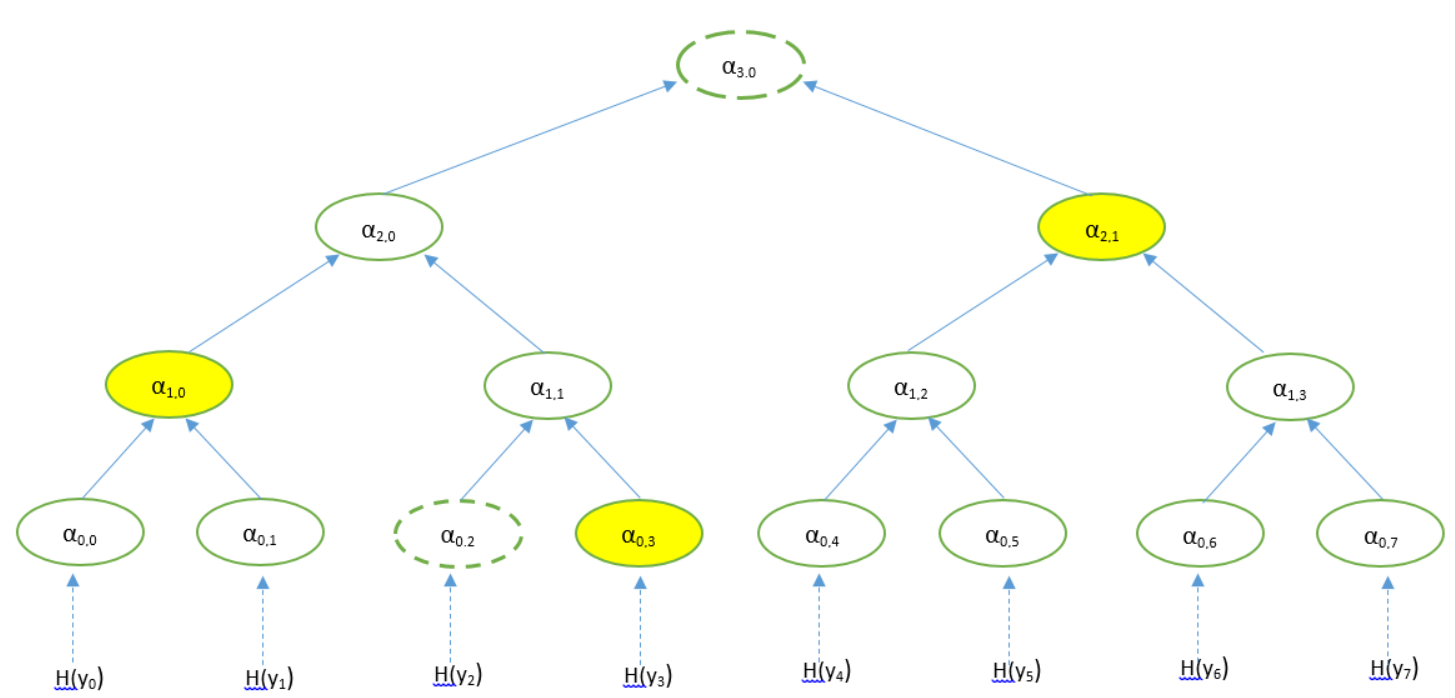}\centering
\caption{An authentication path in a Merkle signature scheme.} 
\label{fig:merkle:signature}
\end{figure}

A Merkle tree may be the basis for deriving hash-based one-time signatures \cite{Merkle}, generalizing the notion of Lamport's signatures \cite{Lamport}. Each $y_{i}$ corresponds to a private key that is being used to sign a message, whereas the root of the tree is the signer's public key. When such a $y_{i}$ is to be used for signing, the signer signs the message with the pair $(y_{i},H(y_{i}))$  - i.e. the signer reveals the private value $y_{i}$ which allows for the computation of $H(y_{i})$ which in turn is being used for the signature. To allow signature verification, the signer should also reveal the {\em authentication path} of the tree - i.e. all the necessary labels of the intermediate nodes which are needed to verify the root of the tree, that is the signer's known public key. For example, for the case of Merkle tree illustrated in Fig. \ref{fig:merkle}, the verification path for the signature corresponding to $y_{2}$ consists of the labels $a_{0,3}$, $a_{1,0}$ and $a_{2,1}$ (see Fig. \ref{fig:merkle:signature}). The Merkle trees constitute the main building blocks for hash-based post-quantum secure signature schemes, such as  LMS \cite{LMS}, XMSS \cite{Huelsing}, and SPHINCS+ \cite{SPHINCS}. 

\section{The new pseudonymisation technique}
\label{sec:analysis}

In this section, we present how a Merkle tree can be used to generate pseudonyms with the properties {\bf P1} - {\bf P5} described in Section \ref{sec:introduction}. 

\subsection{Initial assumptions}
Our main initial assumption is that an individual $A$ is being identified by different organisations $\mbox{Org}_1,\ldots, \mbox{Org}_N$ though pairwise distinct identifiers $\id_{A_1},\ldots,\id_{A_N}$ respectively; for example, depending on the context, such identifiers could be. e.g., a Social Security Number, an Identity Card Number, a VAT Registration Number or even an e-mail address or a device identifier. We are not interested in how each $\mbox{Org}_{i}$, $i=1,2,\ldots,N$, obtains, at the first place, the identifier $\id_{A_i}$; we simply assume that the validity of each such identifier, for the corresponding organisation, is ensured (e.g. through a secure registration procedure).

Actually, the term {\em identifier} is being used hereinafter to describe any piece of information that allows distinguishing an individual from other individuals; the GDPR (Article $4$) provides a non-exhaustive list of common identifiers (such as name, identification number, location data, online identifier). In our scenario, an identifier for each organisation could be of any form. However, our main  assumption is that an organisation $\mbox{Org}_i$ should not, by default, be able to link personal information of $A$ with another organisation $\mbox{Org}_j$ (i.e. such a linking would vanish the data minimisation principle); this in turn means that, for any pair $i,j$, the organisation $\mbox{Org}_i$ should not know that $\id_{A_j}$ is associated with the individual $A$. For example, a hospital may identify a patient though her Social Security Number; however, since there is no need for the hospital to get knowledge of the VAT Registration Number of the patient, this user's identifier should not become known to the hospital. 
 
It should be pointed out that the notion of identifier is not necessarily restricted to a single value, as in the  examples presented above; for example, one such identifier $\id_{A_i}$ may consist, e.g., of a combination of individual's first name, last name and ID card number; although the ID card number by itself suffices to uniquely identify the individual in a specific context (i.e. it is a single identifier), we may allow incorporating many user's attributes into the definition of a single identifier. By these means, the number of possible single identifiers that may occur does not affect our analysis, since  all of them may become part of a new single identifier incorporating all of them. Indeed, re-visiting the above example, the combination of first name and last name could be also an identifier (depending on the context), apart from the ID card number; however, we may define in this case that:  $\id_{A_1}=\mbox{first\_name}\cnc\mbox{last\_name}\cnc\mbox{ID\_card\_number}$.

Moreover, in the subsequent analysis, we assume that the organisations $\mbox{Org}_i$, $i=1,2,\ldots,N$ do not secretly exchange information towards getting, for a user $A$, much more personal information of $A$ than they need for their data processing purposes (i.e. they do not collaborate maliciously in terms of personal data protection requirements). In other words, for any $i\neq j$,  $\mbox{Org}_{i}$ and  $\mbox{Org}_{j}$  do not try to link information relating to $\id_{A_i}$ and $\id_{A_j}$ - which would also lead in revealing  $\id_{A_i}$ (resp. $\id_{A_j}$) to $\mbox{Org}_{j}$ (resp. $\mbox{Org}_{i}$). It should be noted that such linking attacks,  generally, may be possible between different databases, even if the relevant users identifiers are different: this may occur, for example, by comparing the so-called quasi-identifiers of  individuals (see, e.g., \cite{Fung}). In our paper, we do not consider such threats; we simply focus on deriving user-generated different pseudonyms $P_{A}^{(i)}$ and $P_{A}^{(j)}$ being pairwise unlinkable - i.e., for any $i\neq j$, $P_{A}^{(i)} \neq P_{A}^{(j)}$, whilst knowledge of $P_{A}^{(i)}$, $i=1,2,\ldots,N$, does not allow computation of $\id_{A_i}$, as well as computation of any other identifier $\id_{A_j}$  of $A$. Moreover, under the aforementioned assumption of the honest model with respect to the behavior of the parties (organisations), there is no way to establish that $P_{A}^{(i)}$ and $P_{A}^{(j)}$ correspond to the same individual - unless the individual $A$ by herself wants to prove it. 

\begin{figure}[!t]
\includegraphics[width=1\textwidth]{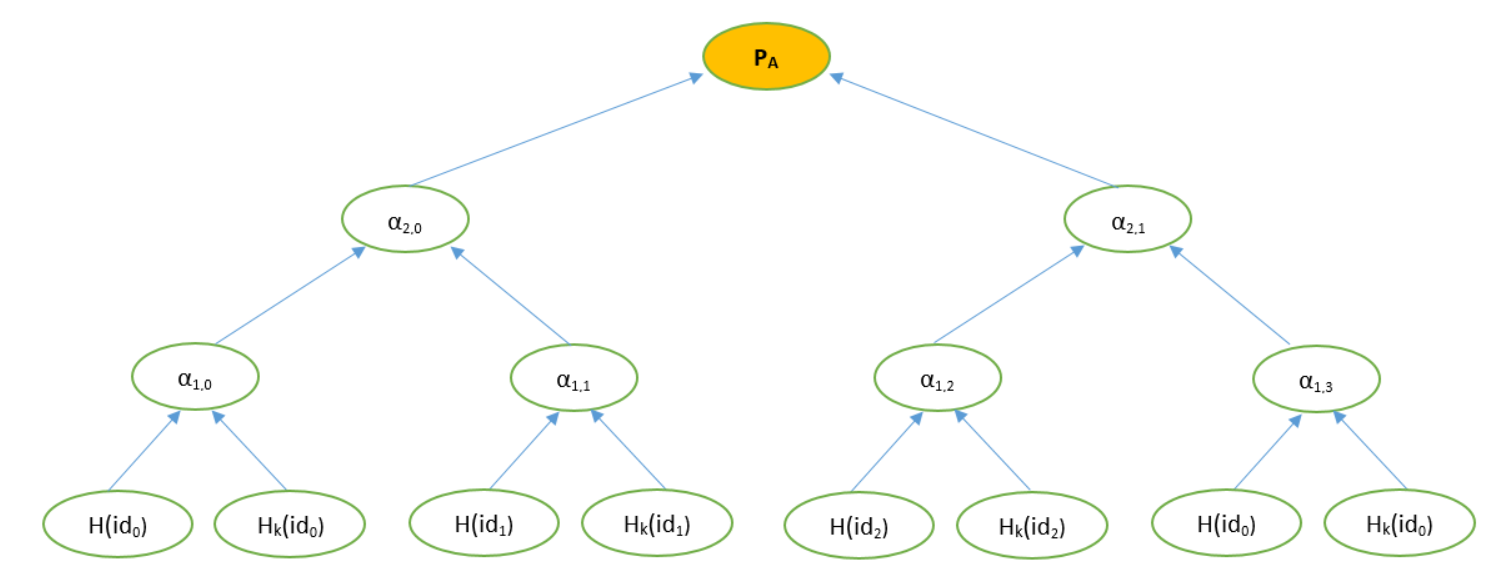}\centering
\caption{Generating a pseudomyn depending on $\id_{A_1},\id_{A_2},\id_{A_3},\id_{A_4}$, (b) Proving that the pseudonym corresponds to the known identifier $\id_{A_2}$}
\label{fig:sub1}
\end{figure}

\begin{figure}[!t]
\includegraphics[width=1\textwidth]{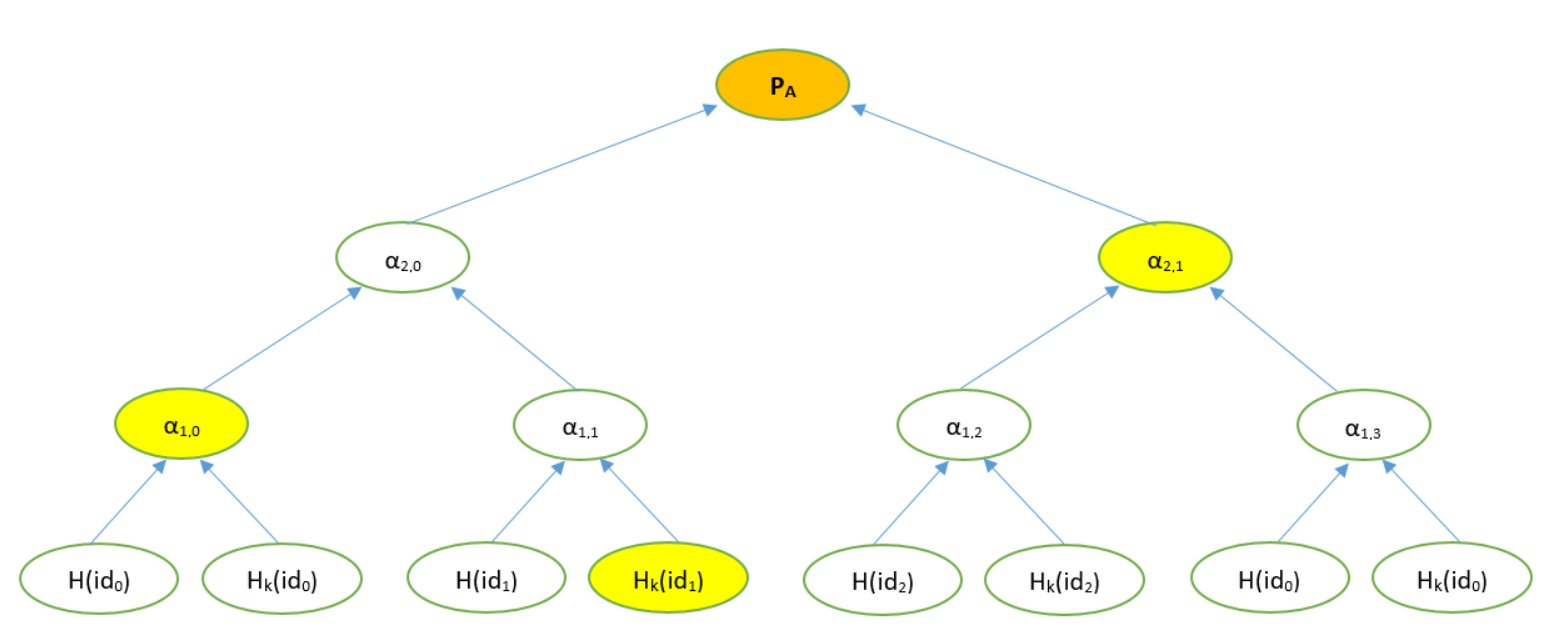}\centering
\caption{Proving that the pseudonym $P_A$ corresponds to the known identifier $\id_{A_1}$}
\label{fig:sub2}
\end{figure}


\subsection{How to derive a pseudonym}

Let us assume that the user $A$ wants to generate $N$ pseudonyms $P_{A}^{(i)}$, $i=1,2,\ldots,N$, each for one different organisation $\mbox{Org}_{i}$, where  $\mbox{Org}_{i}$  already knows the identifier $\id_{A_i}$ of $A$. First, for the sake of simplicity, we assume that $N=2^n$ for some integer $n$ (and the more general case will be subsequently discussed). For these parameters, each pseudonym is associated with a Merkle tree of height $n+1$ - i.e. with $2N$ leaves. For the $i$-th such Merkle tree, the labels $a_{0,0}^{(i)}, a_{0,1}^{(i)},\ldots, a_{0,2N-1}^{(i)}$ of its leaves are constructed via a secret key $k^{(i)}$, which is known only to $A$, as follows:

\begin{equation*}
a_{0,\ell}^{(i)}=\begin{cases}
H(\id_{A_{(\ell+2)/2}}), &\mbox{ if } \ell=0,2,4,\ldots,2N-2,\\
H_{k^{(i)}}(\id_{A_{(\ell+1)/2}}),  &\mbox{ if } \ell=1,3,5,\ldots,2N-1
\end{cases}
\end{equation*}

where $H$ is a cryptographic hash function and $H_{k^{(i)}}$ is a keyed hash function with a key $k^{(i)}$ - i.e. a Message Authentication Code (MAC). Actually, in simple words, each pair of leaves of the Merkle tree corresponds to one users's identifier; the first leaf of the pair is labelled by the hashed value of the identifier, whilst the second leaf is labelled by the keyed hashed value of the identifier (i.e. its MAC). Such a Merkle tree is shown in the  Fig. \ref{fig:sub1}, with $8$ leaves (i.e. $N=4$); the label $P_{A}^{(i)}$ of the root corresponds to the pseudonym of the user. Apparently, the root of the tree depends on all its leaves - namely, on all identifiers $\id_{A_{i}}$, $i=1,2,\ldots,N$. Moreover, by using different key $k^{(i)}$ for each Merkle tree, a different pseudonym $P_{A}^{(i)}$ is being generated for each $\mbox{Org}_i$; all of these pseudonyms $P_{A}^{(i)}$ though, $i=1,2,\ldots,N$, share the same properties (i.e. each of them depends on all $\id_{A_i}$, $i=1,2,\ldots,N$,) and are pairwise unlinkable. 

Since we assume that these pseudonyms are being generated in the user's environment, which in turn means that each secret key $k^{(i)}$ is known only to the pseudonym's owner $A$, it is essential to establish a procedure that the user $A$ proves to an organisation $\mbox{Org}_j$ that she owns the pseudonym $P_{A}^{(i)}$. Recalling our assumption that $\mbox{Org}_j$ already knows $\id_{A_j}$ as an identifier of $A$, the properties of a Merkle tree as a cryptographic primitive for a digital signature (see Fig. \ref{fig:merkle:signature}) suffice to allow such a proof of ownership. Indeed, let us assume that, for the pseudonym derived in the  Fig. \ref{fig:sub1}, that the user $A$ wants to prove to $\mbox{Org}_2$ that she owns this pseudonym.  Then $A$ simply reveals the labels of the authentication path of the tree that allows $\mbox{Org}_2$ verify that the pseudonym $P_{A}^{(i)}$ corresponds to $\id_{A_2}$ - namely, $A$ reveals the values $H_{k^{(i)}}(\id_{A_{2}})$, $a_{1,0}$ and $a_{2,1}$ (recall that $\mbox{Org}_{2}$ can compute $H(\id_{A_2})$). This is illustrated in  Fig. \ref{fig:sub2}. Clearly, a similar procedure may be followed for any identifier $\id_{A_i}$, $i=1,2,\ldots,N$ or, equivalently, to any organisation $\mbox{Org}_i$, $i=1,2,\ldots,N$. Due to the properties of the Merkle tree, no information on other identifiers of $A$ is revealed in this verification procedure. Moreover, for a secure keyed hash function, knowledge of the pair $(\id_{A_i},H_{k^{(i)}}(\id_{A_{i}}))$ does not allow computing $k^{(i)}$. 

An interesting remark is that the above verification procedure holds for any pair $i,j\in\{1,2,\ldots,N\}$ - i.e. $A$ can always prove to $\mbox{Org}_j$ that $P_{A}^{(i)}$ corresponds to $\id_{A_j}$, without revealing to $\mbox{Org}_{j}$ any other information on the remaining identifiers $\id_{A_{\ell}}$, $\ell\in\{1,2,\ldots,N\}\setminus\{j\}$. If $i=j$, such a verification procedure is essential in establishing that the pseudonym of $A$ within $\mbox{Org}_j$ will be $P_{A}^{(j)}$; this is a necessary first step (i.e. a registration process), since a pseudonym $P_{A}^{(j)}$ may be considered as valid in a specific context only if the relevant organisation $\mbox{Org}_{j}$ that will use $P_{A}^{(j)}$ is ensured for its validity. However, if $i\neq j$, then we are in the case that $A$ proves to $\mbox{Org}_{j}$ that she owns another pseudonym $P_{A}^{(i)}$ in another organisation $\mbox{Org}_{i}$ - i.e. the user $A$ allows for linking personal information of herself. Note that, without this intervention (i.e. proof of ownership) of $A$, such a linking between $P_{A}^{(i)}$ and $P_{A}^{(j)}$ is not possible.

Due to the nature of the Merkle tree, which is typically considered as a full balanced binary tree, the number of its leaves is always a power of $2$; that's why we first assumed that $N=2^n$ for some $n$. However, this may not be always the case. To alleviate this issue, we choose the minimum possible integer $m$ such that $N<2^m$ and we proceed accordingly by constructing a Merkle tree with $2^{m+1}$ leaves; the first $2N$ of them correspond to the $N$ user's identifiers as described previously, whereas the remaining $2^{m+1}-2N$ leaves can be chosen arbitrarily.

\subsection{Security requirements}
\label{sub:security:analysis}

As in any hash-based post-quantum signature scheme, the security of the described pseudonymisation scheme  rests with the cryptographic properties of the underlying hash function $H$ in the Merkle tree \cite{Buchmann:b,Buchmann:c}. In our case, the necessary security requirement for $H$ is collision resistance, which means that finding two inputs with the same hash value is computationally infeasible. Collision resistance implies other weaker security requirements such as one-wayness (i.e. for essentially all pre-specified outputs, it is computationally infeasible to find any input which hashes to that output) and second-preimage resistance (i.e. it is computationally infeasible to find a second input which has the same hash value as that of a specified input). Therefore, if collision resistance is present, an adversary cannot find, for any given pseudonym $P_{A}^{(i)}$, as well as for any given label of an intermediate node of the tree, any relevant $\id_{A_{\ell}}$, $\ell=1,2,\ldots,N$. Moreover, an adversary cannot find/compute  inputs which give rise to the same outputs as the original $\id_{A_{\ell}}$, $\ell=1,2,\ldots,N$.

An important parameter of hash functions towards evaluating the aforementioned collision resistance is the length $n$ of the hash value. As it is stated in \cite{Buchmann:b}, the so-called birthday attack, which works for any hash function, finds a collision in time approximately $2^{n/2}$; moreover, in a quantum world, there exists an attack finding a collision in time approximately $2^{n/3}$ \cite{Brassard}. Therefore, today we need $n\geq 256$, whilst for post-quantum security we need $n\geq 384$ \cite{Buchmann:b}. The cryptographic NIST standards SHA-2 \cite{sha2} and SHA-3 \cite{sha3} are known to be collision-resistant, also supporting such lengths for their output.

Our pseudonymisation scheme also utilises a keyed hash function. Apparently, the requirement of collision resistance is also necessary for this function. To this end, the HMAC \cite{hmac} or GMAC \cite{gmac} cryptographic standards  can be, e.g., used. For the case of HMAC, it is well-known \cite{Bellare} that its cryptographic strength depends on the properties of the underlying hash function - whereas the security of HMAC is formally studied in \cite{Bellare:b}. Therefore, again an appropriate collision resistant hash function, such as SHA-2 or SHA-3, suffices to develop a secure keyed hash function (see also ENISA's report on cryptographic parameters \cite{enisa:crypto}). The security properties of the GMAC, being an information-theoretic authentication code, already assumes an attacker with unlimited computing power and, thus, it provides protection in the post-quantum era (see, e.g., \cite{Bernstein}). Regarding the key size, GMAC provides post-quantum security with $128$ bits key size \cite{Bernstein}. The same key size is also considered as adequate today for HMAC; however, although there is a belief that such a key size will also be adequate for HMAC in the post-quantum era, several post-quantum implementations of HMAC utilize larger key sizes (see, e.g., \cite{Latif}).

Apart from the properties of the cryptographic primitives that are being used, the following security requirement should also necessarily be in place: When the user $A$ registers her pseudonym $P_{A}^{(i)}$ to the organisation $\mbox{Org}_{i}$, this registration should be authenticated (i.e. the identity of $A$ providing $\id_{A_i}$ shall be ensured). Otherwise, an adversary having knowledge of the original identifier $\id_{A_i}$ of $A$ could clearly create a fake pseudonym $P_{A}^{(i')}$ for which he could prove that it stems from $\id_{A_i}$.

\subsection{Possible applications}
\label{sub:applications}

Based on the above properties of the pseudonymisation scheme, several possible application scenarios can be considered, in which the proposed pseudonymisation scheme provides the means to ensure data minimisation. Some  indicative applications are given below, actually based on the properties $\bf{P1}-\bf{P5}$ of the scheme.

\subsubsection{Minimizing information requested from another entity/data controller}

Let us assume that the individual $A$ is identified by $\id_{A_1}$ in $\mbox{Org}_1$ and by $\id_{A_2}$ in $\mbox{Org}_2$, whereas the organisation $\mbox{Org}_1$ needs to get some information about $A$ from $\mbox{Org}_2$. However, $\mbox{Org}_1$ (resp. $\mbox{Org}_2$) should not get knowledge of $\id_{A_2}$ (resp. $\mbox{Org}_1$). 

As a specific indicative example, $\mbox{Org}_1$ could be, e.g., a University, using the student number of $A$ as $\id_{A_1}$, whereas $\mbox{Org}_2$ could be the Finance Ministry, with the Department/Service that is responsible for citizens taxes, using the VAT number as $\id_{A_2}$. Let us assume that the University needs to get knowledge on the annual income of the student $A$, in order to decide whether the student $A$ deserves some benefits (e.g. housing alliance). Actually, due to the data minimisation principle, if the decision on students benefits is based on a threshold of their annual income, only knowledge on whether this income is higher or lower than the threshold is needed (and not the exact value of the income).

With our pseudonymisation scheme, the user $A$ has already created the pseudonyms $P_{A}^{(1)}$ and $P_{A}^{(2)}$ for the University and Finance Ministry respectively, having proved to each of them their ownership. Each of these pseudonyms is the root of a Merkle tree with $4$ leaves; the leaves, in both trees, corresponds to the student number and the VAT number. To allow the University obtain the necessary information, the student $A$ may provide to her University her pseudonym $P_{A}^{(2)}$ that uses for the service of the Finance Ministry, proving also that this pseudonym indeed corresponds to her; note that, for such a proof of ownership, $A$ will provide to the University the authentication path of the Merkle tree of $P_{A}^{(2)}$ corresponding to her student number and not to the VAT Number (i.e. a different authentication path from the one that $A$ used to prove the ownership of $P_{A}^{(2)}$ to the Finance Ministry). In other words, the University is able to verify the ownership of $P_{A}^{(2)}$, without getting any information on the VAT number of $A$. Then, the University simply asks the Finance Ministry to response whether the annual income of the individual with the pseudonym $P_{A}^{(2)}$ if higher or lower than the prescribed threshold. By these means, the University receives the minimum possible information that is necessary to perform its tasks.

Clearly, other similar examples as the above may be considered.

\subsubsection{Minimizing exchange of information between joint data controllers or between a data controller and a data processor}

There are also cases in which a specific personal data processing is being somehow shared between different entities - each of them having a specific role in the whole process. Depending on the role, this could be a case of joint controllership (i.e. two or more entities are joint data controllers, that is they  jointly determine the purposes and means of processing) or the case that some entities are data processors (i.e. entities which process personal data on behalf of the controller(s)). In any case, there may be necessary that such a single entity should have restricted access to personal data, in the framework of the data minimisation principle. For example, let us assume that one entity (let's say $\mbox{Org}_2$) may perform analysis on raw  data, so as to derive an output for each individual which in turn will be feed to another entity (let's say $\mbox{Org}_1$). It is probable that, due to the data minimisation principle, $\mbox{Org}_2$ should not be able to re-identify the individuals, whilst $\mbox{Org}_1$, being able to re-identify them, should get access only to the outcome of the processing of  $\mbox{Org}_2$ and not to the whole volume of initial raw data.

As a possible application scenario lying in this case, we may refer to data processing based on the Pay-How-You-Drive model. The main idea of this insurance model is that drivers have to pay a premium based on their driving behaviour and degree of exposure, instead of paying a predetermined fixed price. Such a model poses several privacy risks, due to the fact that the evaluation of the driver's behavior typically necessitates tracking of the driver's routes, collecting and/or extracting detailed personal information (speed, harsh braking/acceleration, visited places, trips frequencies and time schedules, number and duration of possible stops etc.). Hence, the proposed pseudonymisation scheme could possibly alleviate such privacy threats as follows:
\begin{itemize}
    \item The collection of raw data, based on the driver's driving information, is performed by $\mbox{Org}_2$, which in turn performs the whole analysis in order to derive a scoring for the driver (according to the model). The scoring by itself does not allow going backwards to the detailed personal information.
    \item $\mbox{Org}_2$ works on pseudonymised information. For an individual (driver) $A$, $\mbox{Org}_2$ uses a user-generated pseudonym $P_{A}^{(2)}$ based on an identifier $\id_{A_2}$ of $A$ which is known only to $\mbox{Org}_2$. Although it is important that $\id_{A_2}$ is unique for $A$ and suffices to discriminate her from any other user, the value $\id_{A_2}$  by itself should not be able to allow finding the identity of $A$. For example, $\id_{A_2}$ could be a unique identifier generated by the relevant smart application. 
    \item $\mbox{Org}_2$ submits the output of its analysis (i.e. scoring of the driver) to the insurance company $\mbox{Org}_1$, in a pseudonymised form, based on the pseudonym $P_{A}^{(2)}$. Note that, at this moment, $\mbox{Org}_1$ is not able to link $P_{A}^{(2)}$ to any of its insured customers (whereas even $\id_{A_2}$ is not known to $\mbox{Org}_1$). 
    \item $\mbox{Org}_2$ deletes the raw data.
    \item The user $A$ proves to the $\mbox{Org}_1$ that she owns the pseudonym $P_{A}^{(2)}$. Such a proof of ownership is based on an identifier $\id_{A_1}$, which is known only to $\mbox{Org}_1$ for identifying its customers, which had been also used, in conjunction with $\id_{A_2}$, for constructing $P_{A}^{(2)}$. In other words, $P_{A}^{(2)}$ is the root of a Merkle tree, whose four leaves are based on two identifiers  $\id_{A_1}$ and $\id_{A_2}$  - namely, they are being labelled by $\id_{A_1}$, $H_{k}(\id_{A_1})$,$\id_{A_2}$, $H_{k}(\id_{A_2})$ respectively.
\end{itemize}
By the above procedure, the insurance company gets the desired information on the evaluation of the driving behavior of its customer $A$, without obtaining any detailed personal information of her driving habits (trips, speed etc.). Moreover, $\mbox{Org}_2$ which get (and processes) such detailed personal information, is not able to link this information to an identified individual, unless of course the user $A$ wants to prove her exact identity to $\mbox{Org}_2$ - e.g. in case of disputing the accuracy of personal data processed (recall also our assumption with respect to the honest model, in the beginning of Section \ref{sec:analysis}). 

Although the above clearly does not constitute a full solution to the privacy challenges of a Pay-How-You-Drive model, it becomes clear that by utilizing  advanced pseudonymisation techniques such the one presented here, more options for alleviating data protection issues in several application scenarios are present.

\subsection{Implementation analysis}

The performance of Merkle trees in terms of one-time signature schemes, taking into account both speed and storage, has been widely discussed by several researchers, whereas several improvements deviating from the naive implementation have been also proposed. Namely, there are techniques aiming to generate the signature without saving too many nodes, at a still efficient time - i.e. dealing with the so-called Merkle tree traversal problem (see, e.g., \cite{Berman,Merkle:a,Szydlo}).

Here, we do not focus on finding the best  possible implementation, but we rather discuss several aspects that need to be considered. First, the height of the tree depends fully on the number of leaves - which, in our case, is fully determined by the number $N$ of original user's identifiers $\id_{A_i}$, $i=1,2,\ldots,N$. The height in turn determines the total number of the tree's nodes, as well as the length of the verification path (which in general affects network performance). More precisely, in a Merkle tree of height $h$, the verification path also includes $h$ nodes. For example, if $h=4$, which means that the number of user's domain-specific identifiers is $N=8$, and the size of the hash value is $384$ bits (i.e. to be consistent with the requirements for post-quantum security), then the size of the verification path, to verify the pseudonym of length $384$ bits, is $192$ Bytes.

Moreover, as stated above, since the typical form of a Merkle tree implies a balanced tree (i.e. all leaves reside  at the bottom level $h$, where $h$ is the height of the tree), the number of the leaves should be a power of $2$, and this forms a restriction of the whole process (it is already discussed that utilizing dummy values could be a way to deal with this issue). Alternatively, this could be alleviated  by extending the notion of the Merkle tree so as to be imbalanced (see, e.g., \cite{Kandappu}); in such a scenario, no all verification paths will have the same length. In any case though, adding a new leaf (i.e. adding a new identifier) yields a new root of the Merkle tree - i.e. a new pseudonym - and, therefore, an extended verification path. We state the above as possible approaches that could be considered as future research steps.

In any case, it is expected that the utilization of a Merkle tree as a vehicle for pseudonymisation through the proposed approach will not lead to a large tree, since typically the number of individual's identifiers will not be too large (see, e.g., the discussion of possible application areas in Subsection \ref{sub:applications}). To verify the effectiveness of the approach for small sizes of  Merkle trees for $N$ identifiers or, equivalently $2N$ leaves, $2\leq N\leq 128$, we executed experiments in a typical Windows 10 home computer (AMD Ryzen $3$ 2200U with Radeon Vega Mobile Gfx $2.5$ GHz, with $8$ Gb RAM). We utilized Python v.$3.8.2$ for our implementation of the pseudonymisation scheme, whilst we measured both the time for creating the pseudonym (i.e. to create the Merkle tree) as well as the time for the pseudonym verification (i.e. to verify that the given verification path does yield the pseudonym, for the given identifier). We utilized the SHA-2 hash function, with hash length $256$ bits, and the HMAC, via the relevant Python libraries. The results illustrate that, for $N\leq 128$, the time for creating a Merkle tree (without implementing the most effective approach) is less than 1 sec (starting from about 15mec for small values of $N$), whereas the time for pseudonym verification is less than 40msec.



\section{Conclusions}
\label{sec:conclusions}

In this paper, we illustrated how a Merkle tree (as a characteristic example of cryptographic accumulator) can be used to provide a pseudonymisation mechanism with specific nice properties in terms of fulfilling the data minimisation principle, which are not easily attained by other known pseudonymisation techniques. Moreover, since the Merkle tree is known to provide post-quantum security, the proposed pseudonymisation scheme suffices to provide long-term security, whereas other known cryptography-based pseudonymisation techniques with similar properties do not provide such post-quantum resistance. The main outcome of the above analysis is that advanced cryptography suffices to provide solutions to personal data protection matters  - not only from a security perspective (which is the obvious one) but also from a data minimisation perspective. 

The ideas presented in this paper opens several directions for further research. First, for any specific case study utilizing this pseudonymisation scheme, a formal security analysis of the whole procedure is essential to be performed. Furthermore, other extensions of the Merkle trees (as in the cases of post-quantum one time signatures like XMSS or SPHINCS) could be possibly studied in the framework of deriving pseudonymisation schemes. Another interesting direction is to examine how the pseudonym's owner will be able to explicitly define which types of personal data will be exchanged between the two entities (organisations), so as to eliminate the risk that the two organisations exchange more personal data than it is necessary (i.e. to force transparency on the data exchange to the user). 

In any case, a general conclusion is that the properties achieved by the proposed technique should be further elaborated, in terms of identifying other application areas in which this technique will be beneficial.

%
%
%

\end{document}